# Chapter 1
# Water Contribution to the Protein Folding and its Relevance in Protein Design and Protein Aggregation

**Giancarlo Franzese[1], Joan Àguila Rojas, Valentino Bianco, Ivan Coluzza**

**Abstract** Water plays a fundamental role in protein stability. However, the effect of the properties of water on the behaviour of proteins is only partially understood. Several theories have been proposed to give insight into the mechanisms of cold and pressure denaturation, or the limits of temperature and pressure above which no protein has a stable, functional state, or how unfolding and aggregation are related. Here we review our results based on a theoretical approach that can rationalise the water contribution to protein solutions' free energy. We show, using Monte Carlo simulations, how we can rationalise experimental data with our recent results. We discuss how our findings can help develop new strategies for the design of novel synthetic biopolymers or possible approaches for mitigating neurodegenerative pathologies.

[1]Giancarlo Franzese (☐) · Joan Àguila Rojas
Secció de Física Estadística i Interdisciplinària–Departament de Física de la Matèria Condensada, Facultat de Física, Universitat de Barcelona, Martí i Franquès 1, Barcelona 08028, Spain
Institute of Nanoscience and Nanotechnology (IN2UB), Universitat de Barcelona, Martí i Franquès 1, Barcelona 08028, Spain
E-mail: gfranzese@ub.edu

Valentino Bianco
Faculty of Chemistry, Chemical Physics Department, Universidad Complutense de Madrid, Plaza de las Ciencias, Ciudad Universitaria, Madrid 28040, Spain
E-mail: vabianco@ucm.edu

Ivan Coluzza
Center for Cooperative Research in Biomaterials (CIC biomaGUNE), Basque Research and Technology Alliance (BRTA), Paseo de Miramon 182, 20014, Donostia San Sebastián, Spain and IKERBASQUE, Basque Foundation for Science, 48013 Bilbao, Spain.
E-mail: icoluzza@cicbiomagune.es

## 1.1 Introduction

Proteins are complex molecules consisting of one or more amino acids chains and are essential for all living organisms. They can be part of structural components, as in muscles and collagen, or can develop fundamental functions, like antibodies and enzymes. To perform their biological activity, proteins in many cases need to fold into unique conformations (native states, one for each protein) and be soluble. Protein misfolding can lead to their aggregation with fibrillar morphology (amyloids), which, as a consequence, gives rise to pathological conditions. These conditions include neurodegenerative disorders, as Alzheimer's or Parkinson's disease, and others amyloidosis, i.e., pathological states associated with the formation of extracellular amyloid deposits [1]. Among them, two pathologies of high interest are organ-limited amyloidosis, e.g., the senile cardiac amyloidosis that causes heart failure, and systemic amyloidoses, e.g., the AL amyloidoses affecting kidneys, heart, peripheral nervous system, gastrointestinal tract, blood, lungs, and skin, among other organs.

A folded protein undergoes a conformational change (denaturation) and unfolds when its temperature $T$ is increased above some limiting value. The unfolding by heating stops the protein's biological function and changes its properties, including optical qualities and solubility. A typical example occurs when we cook an egg. Heat makes the transparent and liquid albumen, a protein solution made of about 90% water and 10% proteins, opaque-white and gelatinous.

The thermal denaturation is well understood as a consequence of the increase of the solution's kinetic energy, which induces disorder into the ordered native state. Hence, hot unfolding is a process somehow similar to the melting of a crystal. The transition from the high-$T$ unfolded state to the lower-$T$ folded configuration is understood as a consequence of the protein chain collapse due to the effective attraction among the amino acid residues [2].

However, proteins unfold also when they are pressurised [3]. Unlike the case of high temperature, the effect of pressure on the stability of the native state is more debated, with some emphasis on the roles of cavities in the folded protein [4].

More intriguing is that proteins can also unfold by decreasing $T$ [5]. Cold denaturation is counterintuitive if we assume that folding is similar to melting, and the protein stability is controlled by an effective attraction that increases when $T$ decreases [6,7]. It reveals that protein stability results from a balance among different water-mediated forces, including, but not limited to, effective hydrophobic interaction.

To rationalise all the experimental data, Hawley proposed a theory that, based on free energy relations, predicts for any protein with a native state a close stability region (SR) with, in general, an elliptic shape in the $T$-$P$ plane. Depending on the specific protein and its measurements, the elliptic SR can also extend toward negative pressures. Hence, the theory predicts that a protein would denature under tension, consistent with experimental extrapolations [8] and direct observations [9].

Cold and $P$ unfolding can be rationalised, assuming an enthalpic gain of the solvent upon the denaturation process without clarifying this gain's origin and its relation with molecular interactions [10]. Alternative theories have hypothesised a $P$-dependent hydrophobic collapse [11] or enthalpic gain and entropic cost [12]. However, none of these approaches can reproduce the entire elliptic SR as Hawley predicted and supported by experiments. As we discuss in this short review, cold and $P$ unfolding can be better understood as a competition between different free energy contributions coming from water, one from hydration water and another from bulk water, reproducing the whole elliptic SR.

Also, within our approach, we rationalise why proteins, stable at ambient conditions, do not necessarily work at extreme $T$-$P$ conditions and why there are $T$-$P$ above which no protein is stable in its native state. By explicitly including water, we can show its evolutionary driving force toward thermophilic proteins with higher surface hydrophilicity at high $T$ and ice-binding proteins with lower surface hydrophilicity at low $T$. On the other hand, the proteins have smaller segregation between the hydrophilic surface and the hydrophobic core if they are stable at high $P$. As we discuss in the following, our theory allows us to introduce a new protein design protocol useful for engineering proteins and drugs working far from ambient conditions.

Finally, we discuss the protein interface effect on the water fluctuations in the protein hydration shell and their relevance in the protein-protein interaction. We study the folding and aggregation of proteins as a function of their concentration. The mechanisms leading to the folding process's failure and the formation of potentially dangerous protein aggregates are a matter of considerable scientific debate [13]. We show that the propensity to aggregate is not strictly related to the surface hydrophobicity of the protein. The increase of the concentration of individual protein species can induce a partial unfolding of the native conformation without the occurrence of aggregates. The concentration at which the proteins aggregate is, indeed, higher than that at which they unfold. Hence, we discuss two-steps smooth transition between the folded, unfolded, and aggregated states of proteins and how this process could be affected by a nearby hydrophobic interface, such as that of a nanoparticle.

This chapter does not pretend to represent an exhaustive review of the vast literature about protein folding, aggregation, and design generated in many decades by the community of renowned biophysicists. It is just a summary of a part of our contributions to the field. It can be considered as a structured introduction to the topic for beginners in this field. It assumes the knowledge of basic notions of Statistical Physics and Thermodynamics and provides a minimal list of references where the interested reader can find further details and more comprehensive reviews.

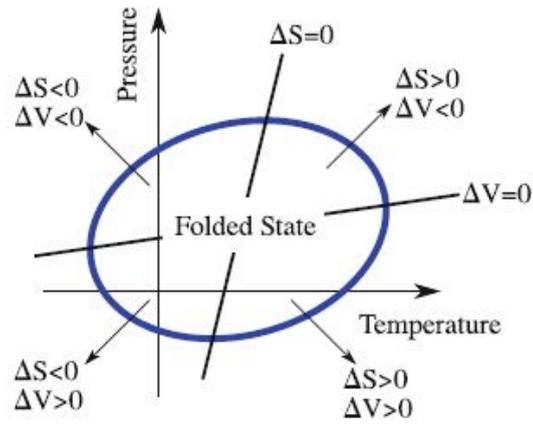

**Fig. 1.1** Stability Region of a protein according to the Hawley theory in a *P-T* representation. The elliptic line separates the native (folded) from the denatured state. This transition occurs by varying either the volume *V* or the entropy *S* or both, as illustrated by the straight lines and arrows. Loci with *ΔV=0* and *ΔS=0* cross the ellipsis where its tangent has infinite or zero slope, respectively.

## 1.2 Hawley's theory

In 1971, Hawley proposed a theory based on the assumption that the folding (f) unfolding (u) transition can be modelled as a reversible first-order phase transition and that equilibrium thermodynamics hold during the denaturation [14]. The theory is based on a Taylor expansion of the difference in Gibbs' free energy $\Delta G$ between the folded and the unfolded state truncated at the second order:

$$\Delta G\left(P,T\right) = \frac{\Delta\beta}{2}\left(P-P_0\right)^2 + 2\Delta\alpha\left(P-P_0\right)\left(T-T_0\right) +$$

$$-\frac{\Delta C_p}{2T_0}(T-T_0)^2 + \Delta V_0\,(P-P_0) - \Delta S_0\,(T-T_0) + \Delta G_0 \qquad (1.1)$$

where $T_0$ and $P_0$ are the temperature and pressure of the ambient conditions, $\Delta V_0$ and $\Delta S_0$ are the volume and entropy variation at the transition, $\alpha$ is the thermal expansivity factor, $C_P$ is the isobaric heat capacity and $\beta$ is the isothermal compressibility factor. This equation is constrained by $\Delta\alpha > \Delta C_p \Delta\beta/T_0$ resulting in the SR having an elliptic shape (Fig. 1.1). Even if it's a phenomenological model, its ability to predict the processes observed in experiments makes it a useful tool when studying these types of systems.

## 1.3 The model for the hydrated protein

We consider a single protein, represented by a flexible chain of coarse-grained residues ($R$), suspended in water ($w$) and described by the enthalpy

$$H \equiv H_{R,R} + H_{R,w} + H_{w,w}^{(h)} + H_{w,w}^{(b)} \qquad (1.2)$$

where the first term accounts for the residue-residue contribution, the second for the residue-water contribution at the protein interface, the third for the hydration ($h$) water contribution, and the fourth for the bulk ($b$) water contribution away from the protein interface. Before describing in details each of these terms, we observe that this minimal model comprises:

1. The covalent (peptide) bonds between the amino-terminal and the carboxyl-terminal of consecutive monomers along the chain, which constrains the possible configurational changes of the protein and the total conformational entropy.

2. The interactions between nonconsecutive residues, including a repulsive and an attractive part, represented as a pairwise term, sum of all the $R$-$R$ (van der Waals, hydrogen bond, and electrostatic) contributions. This term is one of the driving forces leading to the native state.

3. The $R$-$w$ van der Waals and hydrogen bonding enthalpy at the protein surface, influencing the protein secondary structure, exposing the hydrophilic residues to water, and burying the hydrophobic residues into the folded protein core.

4. The $w$-$w$ van der Waals and hydrogen bond enthalpy in the hydration layer, influenced by the nearby protein residues.

5. The $w$-$w$ van der Waals and hydrogen bond enthalpy in the bulk, not affected by the protein residues.

By changing the pressure $P$ and the temperature $T$, as we discuss in the following, the enthalpy and the Gibbs free energy of the system change and drive the protein toward folded or unfolded states.

### 1.3.1 The Franzese-Stanley coarse-grained water model

In a diluted protein solution, water represents more than 90% of the system. It is the thermal bath that continuously interchanges energy, entropy, and enthalpy with the protein. This interchange allows the protein to overcome the potential energy barriers, separating metastable states from the native state, within the range of $P$ and $T$ at which the folded protein is stable. Fully atomistic models can explicitly account for the water contribution to the free energy but are limited in size and time by their large computational cost. All atoms molecular dynamics (MD) simulations with explicit water are very expensive due to the large number of water molecules necessary to solvate even small proteins. For example, MD can simulate proteins up to 80 amino acid residues in ~ 11,000 water molecules for a cumulative simulation time of 643 μs on a specialized Anton supercomputer [15]. On more affordable CPU or GPU hardware all atoms MD calculations with specialized algorithms can simulate, e.g., 42-residues-long proteins in ~ 5,000 water molecules for a cumulative simulation time of 52 μs [16] or 200 μs [17], or a 29-residue-long region of a disordered protein in ~ 8,200 water molecules for an aggregated simulation time of ~ 1.4 ms [18].

If, instead, the model has an implicit solvent with effective $T$ and $P$-dependent force-fields, larger proteins, up to 92 amino acids, can be simulated for milliseconds. However, the drawback, in this case, is that the water effect on the residues interactions is included only in an effective way, losing the transferability of the model to others $T$ and $P$ and changing the dynamics. Possible alternatives combine MD with experimental constraints reaching predictions of the structures of proteins up to 326 amino acids [19] and competing with structural bioinformatics methods [20]. However, also in these cases, it is hard to tell if the protein-folding kinetics resembles the experiments. More details can be found in recent reviews [21], [22].

Another common option is to coarse-grain groups of atoms into larger rigid units and to combine this grouping with implicit water or pseudo-explicit water (charged Lennerad-Jones particles replacing groups of water molecules) [23]. These models gain speed in simulations by reducing the resolution of the description. Hence, their use is limited to those cases in which no atomic resolution is needed and the principal aim is, in general, a qualitative understanding of a specific process. However, these approaches, although useful in several studies, are unable to account for the most specific contribution of water, the hydrogen bonds.

To explicitly include the water into a coarse-grain protein representation, we developed in these years a water model in which we retain the description of the water hydrogen bonds at the molecular level but lose the molecular resolution of the water position, replacing it with a density field. Furthermore, we account in the model for the many-body contribution of the hydrogen bonds [24-29]. We have shown that this contribution is relevant to better describe the water anomalous properties at least on a qualitative level [30-39]. Furthermore, the simple formulation of the model allows us to perform approximated analytic calculations that facilitate the understanding of the relevant mechanisms underlying the properties of the proteins [40-44].

The large reduction of water degrees of freedom that we attain with our specific coarse-graining allows us to study large systems. In particular, we focused on the water under confinement or at hydrophobic interfaces [45-52] or the interface with proteins [53, 54], see also [55-59]. In the following, we will define the model focusing on the protein folding problem [60-64], the protein design [65, 66], and the protein aggregation [67, 68].

In this approach, we partition the bulk into $N$ cells, one per molecule, and consider that, at high $T$ and ambient $P$, when there are no hydrogen bonds, the molecules are homogeneously distributed in volume $V$, each with a proper volume $v=V/N$. We consider the system at constant $N$, $P$, $T$ with a fluctuating volume $V_{TOT}^{(b)}$ and an enthalpy [28, 29]

$$H_{w,w}^{(b)} \equiv \sum_{ij} U\left(r_{ij}\right) - J N_{HB}^{(b)} - J_\sigma N_{coop} + P V_{TOT}^{(b)} \qquad (1.3)$$

where the first term

$$\sum_{ij} U\left(r_{ij}\right) \equiv 4\epsilon \sum_{ij} \left[ \left(\frac{r_0}{r_{ij}}\right)^{12} - \left(\frac{r_0}{r_{ij}}\right)^6 \right] \qquad (1.4)$$

is the Lennard-Jones (LJ) potential energy (isotropic term), being $\epsilon = \frac{5.8kJ}{mol}$ the depth of the potential well, $r_0 = 2.9$ Å the distance from the particle to the well (molecule's hard-core), and $r_{ij}$ the O-O distance. A cut-off of the LJ potential was established at $r = 6 r_0$ to accelerate the calculations.

The second term accounts for the directional contribution of the hydrogen bond (HB) interaction, with $J=1.2\epsilon$ being the energetic gain for each HB. Here,

$$N_{HB}^{(b)} \equiv \sum_{\langle ij \rangle} n_i n_j \delta_{\sigma_{ij}\sigma_{ji}} \qquad (1.5)$$

is the number of HBs in the bulk, depending on the local density index $n_i$ and the bonding index $\sigma_{ij}$, with $\delta_{\sigma_{ij}\sigma_{ji}} =1$ if molecules $i$ and $j$ have the same bonding index, $\delta_{\sigma_{ij}\sigma_{ji}} =0$ otherwise. The first index is $n_i = 1$ if the proper volume of the molecule $i$ has a characteristic size at most ~3.7 Å, the distance at which a HB breaks [28, 29], $n_i = 0$ otherwise. The bonding index $\sigma_{ij} = 1, \ldots, q$ describes the relative orientation of molecules $i$ to its neighbor $j$. We choose the parameter $q$ by selecting 30° as the maximum deviation from a linear bond (i.e., $q \equiv 180°/30° = 6$) [28, 29]. The sum in (1.5) runs over neighboring molecules and, to simplify the model, we assume that each molecule can form at most four HBs.

The third term of Eq.(1.3) takes into account the quantum many-body interaction caused when a new HB is formed, reinforcing all the other HBs formed by the same molecule [28, 29]. It is

$$N_{coop} \equiv \sum_i n_i \sum_{(l,k)_i} \delta_{\sigma_{ik}\sigma_{il}} \qquad (1.6)$$

and it mimics the cooperativity of the HBs between the possible pairs of the $\sigma_{ij}$ indices of the molecule $i$. The choice $J_\sigma = 0.2\epsilon$, an order of magnitude smaller than $J$, ensures that the cooperative term is relevant only below the temperature at which the HBs are formed.

In Eq.(1.3) the total volume is [28, 29]

$$V_{TOT}^{(b)} \equiv V + N_{HB}^{(b)} v_{HB}^{(b)} \qquad (1.7)$$

where $v_{HB}^{(b)}/v_0 = 0.5$, the average volume increase associated with the formation of a HB, is given by the average volume increase between high-density ices VI and VIII and low-density (tetrahedral) ice Ih, and $v_0$ is the proper volume of a water molecule without the HB. The volume term $V$ is free to fluctuate as a function of the LJ interaction and depends weekly on $T$ and $P$ near ambient conditions. The local formation of HBs gives rise to the local density fluctuations responsible for the density and compressibility anomaly of water [30-39].

### 1.3.2 The coarse-grained hydration water model

As we describe in the following, we coarse-grained the proteins as self-avoiding heteropolymers with residue that can occupy only one of the cells of the system. Hence, when we include proteins in the system, we replace water cells with coarse-grained residues, generating an interface. Each residue can be a neighbor of another residue or can be hydrated. The water can form HBs within the hydration shell. The energy of the hydration-shell HBs depends on the nature of the nearby amino acids. For water molecules, forming a HB, near two hydrophobic ($\Phi$) amino acids, two hydrophilic ($\zeta$) amino acids, or one of each kind (mixed, $\chi$), the enthalpy is [63-65]

$$H_{w,w}^{(h)} \equiv H_{w,w}^{(\Phi)} + H_{w,w}^{(\zeta)} + H_{w,w}^{(\chi)} \qquad (1.8)$$

where

$$H_{w,w}^{(\Phi)} \equiv \sum_{ij} U\left(r_{ij}\right) - J^{(\Phi)} N_{HB}^{(\Phi)} - J_\sigma^{(\Phi)} N_{coop}^{(\Phi)} + P V^{(\Phi)} \qquad (1.9)$$

with

$$V^{(\Phi)} \equiv N_{HB}^{(\Phi)} v_{HB}^{(\Phi)} \qquad (1.10)$$

and

$$v_{HB}^{(\Phi)}/v_{HB,0}^{(\Phi)} \equiv 1 - k_1^{(\Phi)}P \tag{1.11}$$

where $v_{HB,0}^{(\Phi)}$ is the volume increase for $P = 0$, and $k_1^{(\Phi)} = \frac{v_0}{4\epsilon}$ is a factor accounting for the compressibility of the hydrophobic hydration shell [63-65]. Based on numerical and experimental observations that point to a higher correlation in the HBs between water molecules near hydrophobic residues, Bianco et al. [63-65] adopted $J^{(\Phi)} = 1.83\,J$, while $J_\sigma^{(\Phi)} = J_\sigma$.

Expressions similar to Eq.(1.9)-(1.11) hold for the hydrophilic ($\zeta$) and the mixed ($\chi$) cases in Eq.(1.8). Parameters in the mixed case are set equal to the average of the $\Phi$ and the $\zeta$ case. To simplify the model, the parameters for the $\zeta$ case are set equal to the bulk case and $k^{(\zeta)} = 0$ because the compressibility effect is due mainly to the hydrophobic interface [63-65].

The total volume occupied by the water is

$$V_{TOT} \equiv V_{TOT}^{(b)} + V^{(\Phi)} + V^{(\zeta)} + V^{(\chi)} \tag{1.12}$$

and fluctuate with $T$ and $P$. The volume occupied by the proteins is constant.

### 1.3.3 The coarse-grained protein-water model

The protein's enthalpy, including the interaction with the hydration shell, is modeled as [63-65]

$$H_p \equiv H_{R,R} + H_{R,w} = \sum_i^{N_c}\left[\sum_{i \neq j}^{N_c}C_{ij}S_{ij} + \sum_j^{N_w}C_{ij}S_i^W\right] \tag{1.13}$$

where the first sum runs over the $N_c$ residues indices, the second over the $N_w$ hydration water molecules, $C$ is a contact matrix with $C_{mn} = 1$ if $m$ and $n$ are first neighbors and 0 otherwise, $S$ is the Miyazawa-Jernigan matrix [69] where $S_{ij}$ accounts for the interaction between amino acids $i$ and $j$, and $S_i^W$ is the interaction energy the residue $i$ with the hydration water molecule and depends on residue's hydropathy, with $S_i^W = -\epsilon^{(\Phi)}$ or $-\epsilon^{(\zeta)}$ if the residue is hydrophobic or hydrophilic, respectively.

With this model, Bianco and Franzese [63] identify the different mechanisms with which water participates in the cold and pressure denaturation. Furthermore, Bianco et al. in Ref. [64] analyze how changes in i) the specific protein residue-residue interactions in the native state of the amino acids sequence, and ii) the water properties at the hydration interface, affect the protein stability region. They show that the solvent properties are essential to rationalize the stability region shape at low $T$ and high $P$, independent of the model's parameters and for both proteins with a native state and disordered proteins. These results open the perspective to develop advanced computational design tools for protein engineering.

## 1.4 Protein design

Protein design has as main objective to understand how a protein sequence encodes specific structural and functional properties [70]. It consists of searching for the ensemble of sequences that fold into a target protein backbone structure. However, it remains one of the major challenges across the disciplines of biology, physics, and chemistry [71, 72].

Protein design is often referred to as the inverse folding problem (IFP). IFP is still one of the toughest major challenges in biophysics, and the most common approach is to search for the sequence that minimizes the energy or a scoring function into the target protein [70, 73-75]. However, until the work by Bianco et al. [65], the way the design changes at extreme conditions was not considered.

In Ref.[65] the authors develop a strategy for protein design, following the previous approach introduced by Couzza et al [73, 74], and focusing on the relationship between the sequence and the thermodynamic conditions at which the target structure is stable. They adopt the coarse-grained model for hydrated proteins described in the previous section to perform the design of heteropolymers on a lattice.

According to previous design studies on the lattice a solution to the folding problem can be found by looking for the sequence with the lowest energy under the constraint that the composition of the amino acids maximizes the number of letter permutations $N_P = N!/(n_A! n_B! .. n_Y!)$ , where $N$ is the sequence length and $n_x$ is the number of residues corresponding to the letter $x$.

In the presence of explicit water, the solvent interaction must be taken into account during the energy minimization. However, a simple downhill search does not work because it would be too sensitive to the large noise produced by the water configurations. Hence, the best strategy is to average the energy of each sequence on the ensemble of water configurations, getting an effective free energy per sequence $F_S$. Such an approach is currently only feasible in a water model on a lattice, where a large number of solvent configurations per sequence can be scanned efficiently, allowing the design protocol to explore many solutions. The design process consists of finding the sequence that minimizes $F_S$ under the constraint of maximizing $N_P$. As a consequence of this approach, the ensemble of the design solutions is influenced by the environmental parameters, such as temperature and pressure, via their effect on the solvent properties. An interesting extension would be to include also the effect of pH changes on the solvent.

### *1.4.1 Design protocols*

Bianco et al. [65] follow two protocols to design a stable protein at a given $T$ and $P$. Both protocols are based on the average enthalpy associated with the hydrated protein. The first (MIN ENTHALPY) consists of minimizing the average enthalpy in the protein folded state. The second (MAX GAP) maximizes the enthalpy gap between the folded and the unfolded protein conformations.

In the following, we adopt the first protocol. We start from an initial random sequence for the protein already folded in the target structure and we sample the space of sequences. At each sampling step, we replace an amino-acid of the polymer either by exchanging it with another of the 20 possible monomers or by swapping the position of two amino acids in the sequence. We perform 100 replacements for each swap.

For each move, we check (i) if it does not lead to a more homogeneous sequence, that would be unable to fold, and (ii) if it decreases the protein's enthalpy. If both conditions are met, we accept the protein move, otherwise we accept it with probability [65]

$$p_{acc} \equiv min\{1, \ exp(-\Delta/T_0\} min\{1, \ (\Pi_n/\Pi_o)^{\omega}\} \qquad (1.14)$$

where $\Delta$ is the change in enthalpy, expressed in internal units, between the new and the old sequence, $T_0 = 0.05$ is an optimization temperature, $\Pi_n$ and $\Pi_o$ are the numbers of permutations of the new and old sequence, respectively, $\omega = 14$ is a weighting parameter. The interested reader can find further details and references in Ref. [65]. Next, we perform a series of Monte Carlo steps to equilibrate the water with cluster and single variable moves [27, 28]. Once the water is equilibrated, we compute the protein's average enthalpy and check that it is smaller than the previous one. Hence, the new sequence is a better candidate for the protein to fold into the target structure at the given pressure and temperature.

For each thermodynamic state point, we sample more than $10^8$ independent sequences and consider, for characterization and stability analysis, only the best 5 or 15 sequences. For these selected sequences we check their validity (ability to fold into the target structure) with isobaric, isothermal Monte Carlo simulations, starting from a stretched protein conformation. We finally select the valid sequence with the smallest enthalpy in the native state [65].

### 1.4.2 Design results in two dimensions.

In Ref.[65], Bianco et al. identify 5–15 optimized sequences for each design pressure and temperature, for a total of more than $1.5 \times 10^3$ optimizations, a number far beyond the capability of any fully atomistic protein model. For each designed sequence, they test the stability in $T$–$P$ by checking if the protein folds into the target structure.

Consistent with natural proteins, they show that sequences that are stable at ambient conditions do not necessarily work at extreme conditions of $T$ and $P$ and that the range of stability in $T$ and $P$ increases with the design temperature $T_d$ at ambient pressure (Fig.1.2). Also, they find that there are limits of $T$ and $P$ above which no protein has a stable functional state [76-78] and no stable sequence can be found. Furthermore, small proteins undergo cold denaturation for temperatures that are lower than those for longer proteins, with a higher content of hydrophobic residues [65]. They demonstrate that these limits, and the selection mechanisms for proteins, depend on how the properties of the surrounding water change with $T$ and $P$.

They study how the design temperature and pressure, $T_d$ and $P_d$, affect the segregation of a hydrophilic surface and a hydrophobic core in the selected proteins. They find that the segregation is never extreme, but is larger at high $T_d$ and low $P_d$ (Fig.1.3). This result is consistent with experimental trends for thermophilic, mesophilic [79], and ice-binding proteins [65].

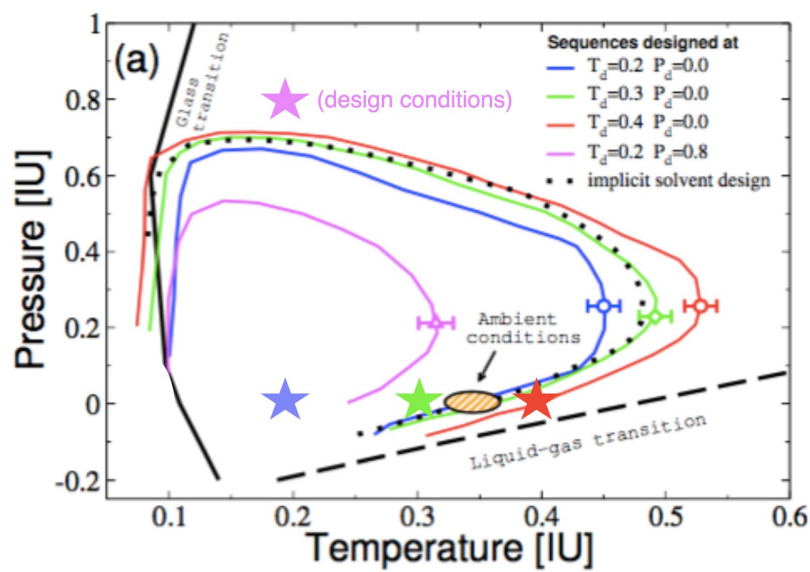

**Fig. 1.2** *T-P* stability region (SR) for proteins designed to fold in a given native structure, with the explicit water model described in the text. The colored continuous lines enclose the SRs for sequences designed at the temperature and pressure indicated in the legend and marked by corresponding colored stars. The dotted line marks the SR found with implicit water. The continuous black line locates the glass transition below which the relaxation times toward equilibrium exceed our observation times (here the Monte Carlo simulation times) as in real experiments. The dashed line traces the liquid-gas spinodal. The orange-shaded ellipse evidences the near-ambient conditions. The symbols with error bars represent the typical error over the estimate of the SRs. Pressure and temperature are expressed in internal units (IU), corresponding to $4\,\varepsilon/v_0$ and $4\,\varepsilon/k_B$, respectively, where $k_B$ is the Boltzmann constant. Adapted from Ref. [65].

Surprisingly, they observe that larger segregation reduces the stability range in $T$ and $P$. The more segregated sequences, selected at high T, are highly resistant (*superstable*) to both cold and pressure denaturation. Superstable proteins have an average stability region that encompasses the average stability region of proteins designed at ambient conditions. These results are potentially useful for engineering proteins and drugs working far from ambient conditions and offer an alternative rationale to the evolutionary action exerted by the environment in extreme conditions [65].

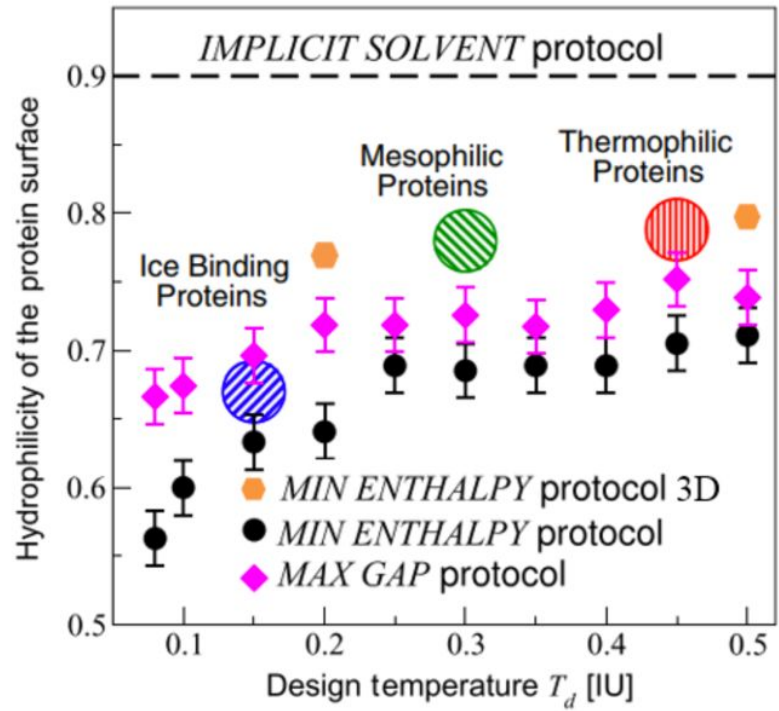

**Fig. 1.3** Hydrophilicity of the surface of the designed protein for a range of design temperatures $T_d$ at $P_d=0$ calculated with explicit water. Sequences selected with different protocols in 2D (black dots and purple rhombuses) qualitatively follow the trend observed for the average hydropathy of real thermophilic (large red circle), mesophilic (large green circle), and ice-binding proteins (large blue circle). Differences between the two protocols are discussed in Ref.[65]. Preliminary results in 3D and two different design temperatures (orange hexagons) show a consistent trend. The discontinuous line corresponds to the calculations with implicit water. The design temperature $T_d$ is expressed in internal units (IU), as in Fig. 1.2, and the hydrophilicity as a dimensionless number normalized to 1, corresponding to the fraction of hydrophilic surface residues. Adapted from Ref. [65].

### *1.4.3 Design results in three dimensions.*

Here we describe our preliminary results for the extension of the design approach, described above, to the 3D case. We aim to assess how the dimensionality affects them when compared with those achieved in 2D.

The change of dimensionality implies some differences in the relative weight of the enthalpy and energy terms described in section 1.3. In particular, if we consider a partition of the 3D volume into cubic cells, each molecule (or residue) has six neighbors, instead of four as in the 2D partition into squares. On the one hand, each water molecule can form only up to four HBs, hence the entropy associated with the HB formation largely increases in 3D compared to 2D. On the other hand, the six neighbors in 3D give rise to more terms in the energy contributions to the total enthalpy, partially balancing the increased entropy.

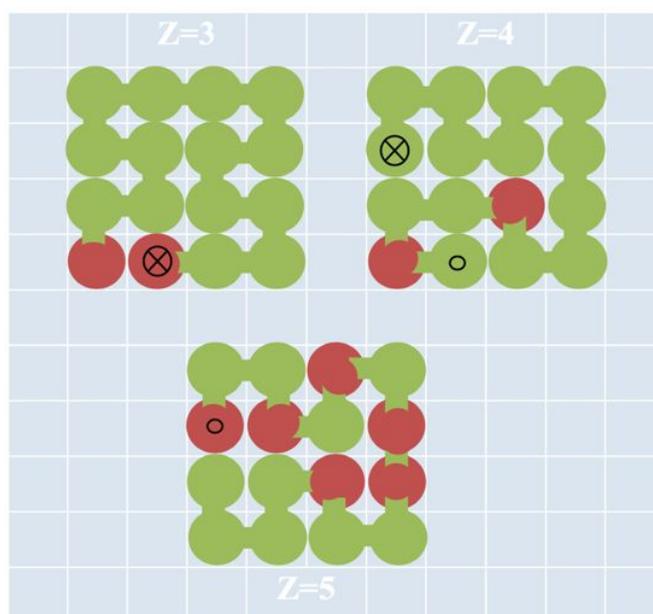

**Fig. 1.4** Schematic representation of the sequence selected by our design protocol at *P=0.8* (4 $\varepsilon/v_0$) and *T=0.2* (4 $\varepsilon/k_B$) for a 48-residue-long heteropolymer with the compact target structure as in the figure. The structure occupies three planes in the cubic partition of the volume, corresponding to the *z*-coordinate *z=3, 4, 5*, with $1 \leq z \leq 9$. Each plane, and the corresponding *z* value, are represented separately for clarity. We mark the chain structure by green segments within the plane and by a crossed or an empty circle—depending on if the chain moves up or down, respectively—between the *z*-planes. We represent the hydrophilic and the hydrophobic residues in green and brown, respectively, according to the Kyte-Doolittle hydropathy scale [80]. All the residues, but the (central) core four at *z=4*, are exposed to water.

To keep our preliminary calculations affordable on cheap GPUs, we design a 48-residue-long heteropolymer with a compact target structure (Fig.1.4). We hydrate the protein with 681 water molecules within a cubic volume partitioned into $9^3$ cells with periodic boundary conditions. The target structure has a core of four residues and a surface of 44 amino acids on three different layers of the partition. We optimize the sequence using the MIN ENTHALPY protocol at three thermodynamic conditions, listed in Table I. For each thermodynamic condition, we observe that the design protocol leads to sequences with converging enthalpies within $10^8$ iterations (Fig.1.5).

**Table 1:** Design pressure and temperature at which we optimize the protein sequence with the target structure in Fig.1.4. Initial and final enthalpy refers to the calculated values at the beginning and the end of the optimization protocol. All the quantities are expressed in internal units (IU): $4\,\varepsilon/v_0$, $4\,\varepsilon/k_B$, $4\,\varepsilon$ for pressure, temperature, and enthalpy, respectively.

| Pressure [IU] | Temperature [IU] | Initial enthalpy [IU] | Final enthalpy [IU] |
|---|---|---|---|
| 0.2 | 0.2 | -0.2 | -12 |
| 0.2 | 0.5 | -0.2 | 14.12 |
| 0.8 | 0.2 | -0.2 | -33.71 |

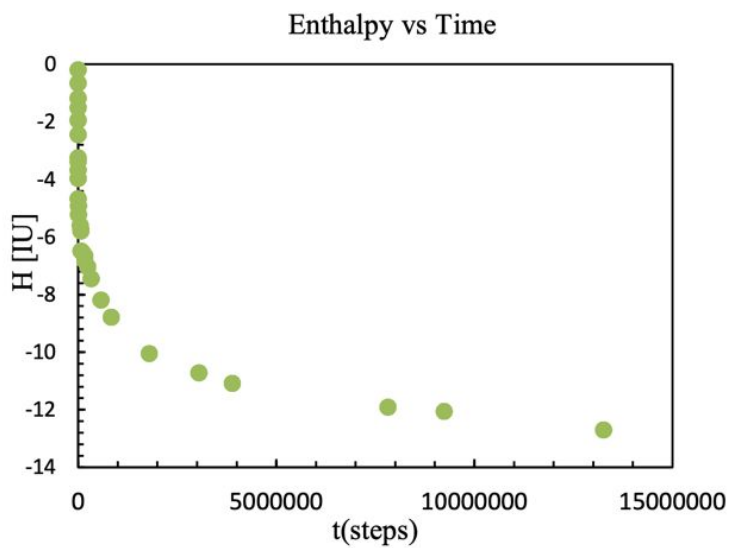

**Fig. 1.5** Enthalpy changes as a function of the number of iterations (time steps) for our design protocol at $P=0.2$ $(4\,\varepsilon/v_0)$ and $T=0.2$ $(4\,\varepsilon/k_B)$ for a 48-residue-long heteropolymer with the compact target structure as in Fig.1.4. Symbols mark the iterations that lead to sequences with lower enthalpy. Enthalpy is measured in internal units (IU), $4\,\varepsilon$.

First, we observe that in 3D (Fig. 1.4) 38, i.e. 80%, of the 44 surface residues are hydrophilic. This value is 30% higher than the surface hydropathy found in 2D at the same $P$ and $T$. However, the 2D result is for a heteropolymer with a smaller number of surface residues, due to its shorter length (30 residues) and to the lower dimensionality. We find that the trend in the difference in surface hydropathy between 2D and 3D design is independent of P and $T$, being always higher in 3D (Table 2).

On the contrary, the fraction of hydrophobic residues in the core in 3D is less than that observed in 2D. Here, only one of the four core residues is hydrophobic, while in 2D approximately 40% of the core was hydrophobic. However, due to the small size of the present core (four residues), any hydropathy smaller than 50% (corresponding to two residues), would be discretized to one single residue.

We rationalize these differences in surface and core hydropathy as due to the different sizes of surface and core sequences. In particular, these results suggest that the smaller the sequence, the smaller the fraction of surface hydrophilic residues or core hydrophobic amino acids. Longer proteins, with larger cores in 3D and larger surfaces in 2D, should be investigated to clarify this point.

Despite these differences between 2D [65] and 3D preliminary results, we find that the $T_d$-dependence of the surface hydropathy in 3D is consistent with that found in 2D and in real thermophilic, mesophilic, and ice-binding proteins (Table 2 and Fig.1.3). In particular, the residue segregation increases by increasing the design temperature $T_d$. More systematic studies, beyond the scope of this minireview, are needed in this direction to support these preliminary results.

**Table 2:** Surface hydropathy (fraction of hydrophilic residues on the protein surface) for the sequences selected by the MIN ENTHALPY protocol in 2D and 3D calculations at different design pressures $P_d$ and temperatures $T_d$. Values in 2D are from Ref.[65]. All quantities are expressed in internal units (IU).

| $P_d$ [IU] | $T_d$ [IU] | 2D Surface Hydropathy | 3D Surface Hydropathy |
|---|---|---|---|
| 0.2 | 0.2 | 0.65 | 0.77 |
| 0.2 | 0.5 | 0.70 | 0.80 |
| 0.8 | 0.2 | 0.55 | 0.80 |

## 1.7 Protein unfolding as a precursor of protein aggregation.

Protein aggregation is a mechanism related to a large number of diseases, such as Spongiform encephalopathies, Parkinson's, or Alzheimer's disease [81-84], and is a relevant issue in biopharmaceutical production [85]. It can be related to the variation of external factors [86] or the milieu composition, as, e.g., ions concentration [87] and it is mostly inevitable when protein concentrations exceed the natural values [67]. Nevertheless, the mechanisms leading to the aggregation are still debated [88] and the objective of extended research. In particular, coarse-grained models represent valid tools to investigate protein solutions at large concentrations in explicit water [67, 89, 90].

In a recent publication, Bianco et al. [67] use the coarse-grain model described in the previous sections to study by Monte Carlo simulations how the increase of concentration of individual protein species affects their folding and aggregation in an aqueous solution. They follow the design protocol described in this chapter to select eight proteins with native structures with maximally compact conformations, composed of 36 or 49 amino acids. The selected proteins have different compositions, with a range of surface and core hydropathy combinations. The fraction of hydrophilic surface amino acids varies from 0.5 to 0.7, while that of hydrophobic residues in the protein core is between 0.25 and 0.45.

They calculate the free energy profile of these proteins as a function of the native contacts and inter-protein contacts, for different protein concentrations $c$. In all the cases they observe that for low concentrations, $c \lesssim 5$, all the proteins reach their native folded state and, on average, are not in contact with each other. They show that the increase of $c$ can induce a partial unfolding of the native conformation without inducing aggregation [67].

At very high (sequence-dependent) concentrations, e.g., $c \gtrsim 20$ in some of the studied cases, Bianco et al. find that the proteins lose completely their folded structures and aggregate. Comparing proteins with different compositions, they show that, as long as the design explicitly accounts for the water environment, the propensity to aggregate is not strictly related to the hydrophobic content of the protein's surfaces. Also, proteins with the same native structure, but different sequences, have thresholds for the unfolding and the aggregation that are different in each case [67].

By approaching two isolated proteins of the same species, they show that their unfolding is water-mediated and starts before the residues can interact directly. This occurs at a distance close to the average separation between proteins at the unfolding concentration. Furthermore, by switching off the water terms between the two approaching proteins, they show that the proteins aggregate without unfolding. Hence, the water causes a free-energy barrier against aggregation [67].

Bianco et al. suggest that the distance at which the two approaching proteins start to unfold can be considered as the water-mediated interaction radius of a protein. To support this conclusion, they perform a percolation analysis of the cluster sizes of statistically correlated water molecules between the two proteins when they are folded or unfolded. This size is a proxy of the correlation length in water affected by the nearby proteins. Bianco et al. find that this size increases when the proteins unfold. Hence, the water-mediated protein-protein interaction increases in the unfolded state. This role of the water in inducing the unfolded regime, which is a precursor of the fully aggregated state, is an unexpected and not previously observed prediction of their simulations [67].

The water-mediated protein-protein interactions are essential also when different proteins are mixed at different concentrations. Bianco et al. [91] show that each component of a protein mixture is capable of maintaining its folded state at densities greater than the one at which they would precipitate in single-species solutions. They demonstrate that the free energy of each protein in the mixture remains unaffected by the presence of the other protein species and depends mainly on their individual concentrations.

Moreover, by increasing the protein concentrations, they find that the protein aggregates mostly between proteins of the same species. The aggregation propensity is essentially regulated by similarities in protein sequence more than in protein structure [91].

They show the generality of their observation over many different proteins, adopting the coarse-grained model of hydrated proteins described here and based on the Franzese-Stanley water model. Furthermore, they confirm their simulation results by performing dynamic light scattering experiments to evaluate the protein aggregation in a binary mixture of bovine serum albumin and consensus tetratricopeptide repeat, clarifying critical aspects of the cellular mechanisms regulating the expression and aggregation of the proteins [91].

## 1.7 Conclusions

The results summarized in this chapter show that water plays a complex role in the enthalpy balance of protein solutions. In particular, protein (i) cold denaturation, (ii) unfolding by pressurization, and (iii) unfolding by depressurization, are energy, density, and enthalpy driven by water, respectively, in addition to other suggested mechanisms [92].

The understanding of these mechanisms and the water contribution to these processes have opened the perspective to develop advanced computational design tools for protein engineering. In particular, including water contribution into the analysis of the selection of stable proteins at ambient and extreme thermodynamic conditions, allows us to elucidate why there are limits of temperature and pressure above which no protein has a stable functional state. Also, it clarifies why the large segregation of hydrophilic amino acids on the surface and hydrophobic residues into the protein core reduces the thermodynamic stability range of a protein.

We presented a design protocol that explains the hydropathy profile of proteins as a consequence of a selection process influenced by water. Proteins selected to be stable at high temperatures are stable also at extremely low temperatures and high pressures and are characterized by non-extreme segregation of a hydrophilic surface and a hydrophobic core. These results shed light on how environmental extreme conditions could have exerted an evolutionary action regulated by the water-mediated interactions.

In particular, the water fluctuations in the hydration shell are essential to understand the protein-protein interactions. The partial unfolding can occur for increasing protein concentration without aggregation. The hydration water is responsible for the enthalpy barrier that stabilizes misfolded proteins when their concentration further increases. As a consequence, the water contribution pushes toward high values the concentration thresholds above which the proteins aggregate, possibly representing an important regulatory mechanism against neurodegenerative aggregation-related diseases.

**Acknowledgments** G.F. acknowledges the support of the Spanish Grant N. PGC2018-099277-B-C22 (MCIU/AEI/ERDF) and the support by ICREA Foundation (ICREA Academia prize). I.C. acknowledges the support of the Maria de Maeztu Units of Excellence Programme – Grant No. MDM-2017-0720 Spanish Ministry of Science, Innovation and Universities, the Austrian Science Fund (FWF) project 26253-N27, the Spanish Ministerio de Economià y Competitividad (MINECO) Grant. N. FIS2017-89471-R, the Programa Red Guipuzcoana de Ciencia, Tecnología y Informacion SN 2019-CIEN-000051-01, the BIKAINTEK program (grant No. 008- B1/2020), the COST Action CA17139.